\documentclass[reprint,aps,prl,amsmath,amssymb,floatfix,nofootinbib]{revtex4-2}

\usepackage{amstext}
\usepackage{graphicx}
\usepackage{esint}

\begin{document}
\title{On the trapped magnetic moment in type-II superconductors}
\author{Ruslan Prozorov}
\email{prozorov@ameslab.gov}

\affiliation{Ames National Laboratory, Ames 50011, IA, USA}
\affiliation{Department of Physics \& Astronomy, Iowa State University, Ames 50011,
USA}
\date{18 June 2024}
\begin{abstract}
Measurements of the trapped (remanent) magnetic moment,  $M_{trap}\left(H\right)$, when a small magnetic field $H$ is turned off after cooling below the superconducting transition temperature, $T_c$, or ramping a magnetic field up and down after cooling in a zero magnetic field, offer substantial advantages in difficult cases of small samples and large field-dependent backgrounds, which is relevant for hydrogen-based ultra-high-$T_{c}$ superconductors (UHTS). Until recently, there was no need for a separate paper on the trapped magnetic flux for well-known critical state models due to the simplicity of the physics involved. However, recent publications showed the need for such an analysis. This note summarizes the expectations for the Bean model with constant critical current density and the Kim model with field-dependent critical currents. It is shown that if the trapped moment is fitted to the power law, $M_{trap}\propto H^{\alpha}$, the fixed exponent $\alpha=2$ is exact for the Bean model, while Kim models show a wide interval of possible values, $2\leq\alpha\leq4$. Furthermore, accounting for reversible magnetization expands the range of possible exponents to $1\leq\alpha\leq4$. In addition, demagnetizing factors are essential and make the trapped moment orientation dependent even in isotropic materials.

As a concrete application, it is shown that flux trapping experiments performed on H$_{3}$S UHTS compounds can be described well using this generalized approach, lending further support to the type-II superconducting nature of H$_{3}$S under ultra-high pressure. 
\end{abstract}
\maketitle

\section{Introduction}

Ultra-high transition temperature, $T_{c}$, hydrogen-based superconductors (UHTS) have attracted enormous attention over the past decade, for their exceptional $T_{c}$, already not too far from room temperature \cite{Pickard2020,Pickett2023,Zhao2023,Troyan2024}. Demonstrating, in principle, that such materials exist is fundamentally important but not easy because of the very high pressure needed. One of the key challenges in proving superconductivity in UHTS is to determine whether their magnetic response is consistent with the expected behavior. The experiments are challenging because small samples, $30-80\:\mu\text{m}$ in diameter and $2-3\:\mu\text{m}$ thick, are inside a massive metallic body of the diamond anvil cell that inevitably has a large background compared to the signal from a tiny sample. As a result, magnetic hysteresis loops, $M\left(H\right)$, are significantly skewed, and background subtraction is not simple, with some degree of uncertainty \cite{Eremets2022,Minkov2022}. Significant effort was put in to verify the results and provide as much experimental detail as possible. Separate papers have been published just to analyze the data already presented \cite{Eremets2022,Prozorov2022}. At the same time, additional experiments were devised. For example, to avoid a large field-induced background, one can measure the trapped magnetic moment. For a more informative experiment, trapped flux can be measured as a function of temperature and of an applied magnetic field that was set before it was switched off.

In case of UHTS, systematic measurements of flux trapping were performed on H$_{3}$S (based) superconductor \cite{Minkov2023}. The trapping experiment was a continuation of a previous work showing diamagnetic screening in H$_{3}$S, although with a substantial background \cite{Minkov2022}. These results were questioned on the grounds that in the Bean model of the critical state, trapped magnetization should follow specific field dependencies depending on thermal and magnetic field history. In particular, for a field-cooled protocol $M_{trap}$ must increase linearly with a magnetic field, $H$, while for a zero-field cooled protocol $M_{trap}\propto H^{2}$ \cite{Hirsch2022}. The published data on H$_3$S were more consistent with the $H-$linear variation in the ZFC case \cite{Minkov2023}. This criticism was questioned theoretically \cite{Budko2024} and was countered experimentally by reproducing similar experimental conditions for small samples of a known type-II superconductor, CaKFe$_{4}$As$_{4}$, inside a bulky pressure cell \cite{Budko2024a}. This work was an extension of a previous extensive study of different small samples of known type-II superconductors, such as pure and Mn-doped CaKFe$_{4}$As$_{4}$ and  MgB$_{2}$ \cite{Budko2023}. An important observation drawn from these experiments was that the magnetic field dependence of the trapped magnetic moment, $M_{trap}\left(H\right)$, depends on the orientation of the applied magnetic field with respect to platelet samples, implying the importance of demagnetizing factors, in addition to crystalline anisotropy. These studies showed qualitatively similar results to the findings in H$_{3}$S \cite{Minkov2023}. This prompted a further critique based on a model that could not fit the $M_{trap}\left(H\right)$ measured in H$_{3}$S, but could be adopted to describe the data in those known superconductors \cite{Hirsch2024,Hirsch2024a}. In a nutshell, the question became: What is the magnetic field dependence of the trapped magnetic moment obtained under different experimental protocols? If we adopt the power-law low-field approximation of a trapped moment, $M_{t}\propto H^{\alpha}$, used to fit the data \cite{Budko2023}, then the Bean critical state model gives for a zero-field cooled protocol a single value $\alpha=2$, whereas a spread of orientation-dependent values $\alpha=1.2-2.0$ was obtained experimentally in known superconductors \cite{Budko2023} and close to $\alpha\simeq1$ in H$_{3}$S \cite{Minkov2023}.  

Due to these discussions and their relevance for the measurements of small samples with a large background, such as in the UHTS hydrides research, it is instructive to summarize the analytic results expected specifically for the remanent magnetization for realistic type-II superconductors. The analysis shows that all the results presented for known superconductors and for H$_{3}$S can be explained within the realistic critical state model.

The critical state model in type-II superconductors was introduced by Bean \cite{Bean1962,Bean1964} and London \cite{London1963}, and in a more general form by Kim \emph{et al.} \cite{Kim1962,Kim1963}. The Kim model was appealing due to the fact that $M\left(H\right)$ derived from it were quite close to experimentally observed, whereas the standard Bean model predicted loops that are almost never realized \cite{Chen1989}. The difference between the two models lies in the field dependence of the critical current density. Due to utmost importance for applied superconductivity, there is vast literature that analyzed the critical state models, calculating all possible static and dynamic irreversible magnetic behaviors. Most likely, the analysis presented in this note has already been discussed, but I am not aware of any publication with explicit derivations.

\section{Remanent (trapped) magnetic moment}

Let $x=0$ be the left boundary of an infinite in the $z-$ and $y-$ directions slab, so the problem is one dimensional. The center is located at $x=d$, and the full width of the slab is $2d$. An external magnetic field $H$ is applied along the $z-$direction. There are three distinct cases of flux trapping to consider: (1) Field cooling experiment (FC), in which the sample is cooled in a magnetic field to a low temperature below the transition temperature $T<T_{c}$. Next, the magnetic field is switched off and the residual (remanent, trapped) magnetic moment is measured. (2) After cooling in a zero field to the base temperature, an external magnetic field, $H$, is applied and then turned off. The field is small enough that the penetrating flux front stops at a distance $x_{p}$, never reaching the center, somewhere in the interval $0\leq x_{p}\leq d$. We label this stage ZFC-1. (3) Same as (2), but magnetic flux penetrates all the way to the center. The trapped magnetic moment will still depend on $H$ until it reaches $H=2H^{*}$, where $H^{*}$ is the characteristic field at which the magnetic flux reaches the center. We label this stage ZFC-2. These general protocols can be considered for any scenario that may include field-dependent current density, reversible magnetization, shape effects, and anisotropies.

Then the strategy is to start with the assumed critical current density and boundary conditions at the edge, figure out the magnetic induction profiles inside the sample, $\mathbf{B}(\mathbf{x})$, and then evaluate the magnetic moment per unit volume using a general equation \cite{Prozorov2018},
\begin{equation}
\mathbf{M}=\frac{1}{V}\int_{V}\left(\frac{\mathbf{B}\left(\mathbf{r}\right)}{\mu_{0}}-\mathbf{H}\right)dV
\label{eq:Mgen}
\end{equation}

In the following, we will absorb the vacuum permeability, $\mu_{0}$, in currents and fields, so that $\mu_{0}j_{c0}\rightarrow j_{c0}$, which leads to a simplified Maxwell equation applicable to our one-dimensional case, $j=-dB/dx$. Noting that in the remanent state, $H=0$, using the slab geometry and mirror symmetry of the magnetic induction with respect to the $(0,y,z)$ plane, Eq.\ref{eq:Mgen} becomes,
\begin{equation}
M_{trap}=\frac{1}{d}\int_{0}^{d}B\left(x\right)dx=\frac{A}{d}\label{eq:Mtrap}
\end{equation}
\noindent where $A$ is the area under the trapped flux profiles in the $H-x$ coordinates.

\begin{figure*}[t]
\includegraphics[width=17cm]{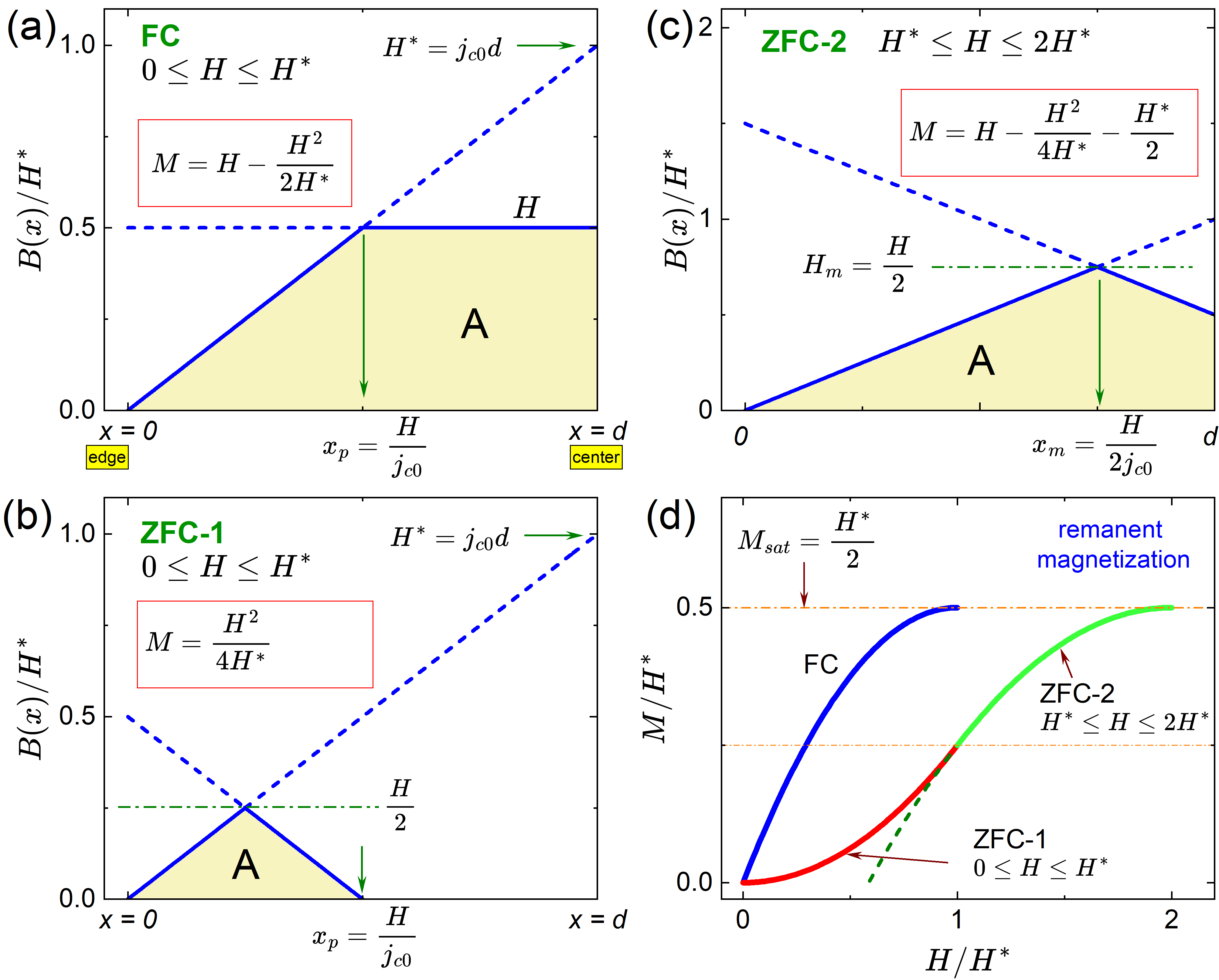} 
\caption{The Bean model with \emph{field-independent} critical current density, $j_{c0}$. Panels (a), (b), and (c) show the magnetic induction profiles in half of the sample for FC, ZFC-1 and ZFC-2 protocols, respectively. Panel (d) shows the resulting field dependence of the trapped volume magnetization by blue, red and green lines, respectively. For convenience, the formulas for the field-dependent remanent magnetization are shown in boxes. Note that magnetization in (d) is normalized by the characteristic full penetration field, $H^{*}=j_{c0}d$. To obtain values normalized by saturation magnetization, $M_{sat}=j_{c0}d/2$, the vertical axis must be multiplied by 2.}
\label{fig:Bean4panels}
\end{figure*}

\subsection{The Bean model}

\subsubsection{Irreversible magnetization only}
\label{subsec:Irreversible-magnetization-only}

In the Bean model, the current density, $j_{c0}$, is independent of a magnetic field. In the FC trapped flux experiment, after the field is switched off, the flux profile at the edge is found from the Maxwell equation, $j=-dB/dx$, setting $j=j_{c0}$. Starting at $x=0$, $B\left(x\right)=j_{c0}x$ and when it reaches the applied field $H$, it levels off at this value. The resulting profile is shown in Fig.\ref{fig:Bean4panels}(a). Evaluating the area and using Eq.\ref{eq:Mtrap}, we obtain the field-dependent trapped magnetization per unit volume, 
\begin{equation}
M_{FC}=H-\frac{H^{2}}{2H^{\ast}},\qquad0\leq H\leq H^{*}\label{eq:BeanFC}
\end{equation}
\noindent where the magnetic field of full flux penetration, $H^{*}=j_{c0}d$. The expression Eq.\ref{eq:BeanFC} is valid in the interval, $0\leq H\leq H^{*}$. Above that field, the trapped magnetic moment saturates at $M_{sat}\left(H\geq H^{*}\right)=H^{*}/2=j_{c0}d/2$. The resulting field-dependent magnetization in shown in Fig.\ref{fig:Bean4panels}(d) by the blue line.

Next, we cool the sample to a low temperature without a magnetic field, and then the magnetic field $H$ is applied. The magnetic induction profile inside the sample is then $B\left(x\right)=H-j_{c0}x$. The flux penetrates the depth $x_{p}$ found from $B\left(x_{p}\right)=0$, or $x_{p}=H/j_{c0}$. When the magnetic field is turned off, it forms the profile where $B\left(x\right)$ decreases to zero at the edge. The resultant triangular trapped magnetic induction profile is shown in Fig.\ref{fig:Bean4panels}(b). Evaluating the area $A$ and using the equation Eq.\ref{eq:Mtrap} we readily obtain, 
\begin{equation}
M_{ZFC-1}=\frac{H^{2}}{4H^{*}}=\frac{H^{2}}{4dj_{c0}},\qquad0\leq H\leq H^{*}\label{eq:Mzfc1}
\end{equation}
This purely quadratic field-dependent ZFC-1 magnetization in shown in Fig.\ref{fig:Bean4panels}(d) by the red line.  

Finally, for completeness, we analyze the ZFC-2 case. Following similar arguments, we find the magnetic induction profile for some intermediate field, $H^{*}\leq H\leq2H^{*}$, shown in Fig.\ref{fig:Bean4panels}(c). Finding the area of the shaded figure, we obtain the following. 
\begin{equation}
M_{ZFC-2}=H-\frac{H^{2}}{4H^{*}}-\frac{H^{*}}{2},\qquad H^{*}\leq H\leq2H^{*}\label{eq:Mzfc2}
\end{equation}
\noindent which, indeed, saturates at $M_{sat}=H^{*}/2$ as the FC magnetization, but in this case it requires the application of twice as large of a magnetic field, $H=2H^{*}$, to reach saturation. The resulting field dependence is shown in Fig.\ref{fig:Bean4panels}(d) by the green line. In summary, Fig.\ref{fig:Bean4panels}(d) shows the trapped field-dependent volume magnetization for the Bean model with magnetic field-independent critical current density, $j_{c0}$, for three distinct experimental protocols. Note that magnetization is normalized by the characteristic full penetration field, $H^{*}=j_{c0}d$. To obtain values normalized by saturation magnetization, $M_{sat}=j_{c0}d/2$, the vertical axis must be multiplied by 2.

\begin{figure}[tb]
\includegraphics[width=8.5cm]{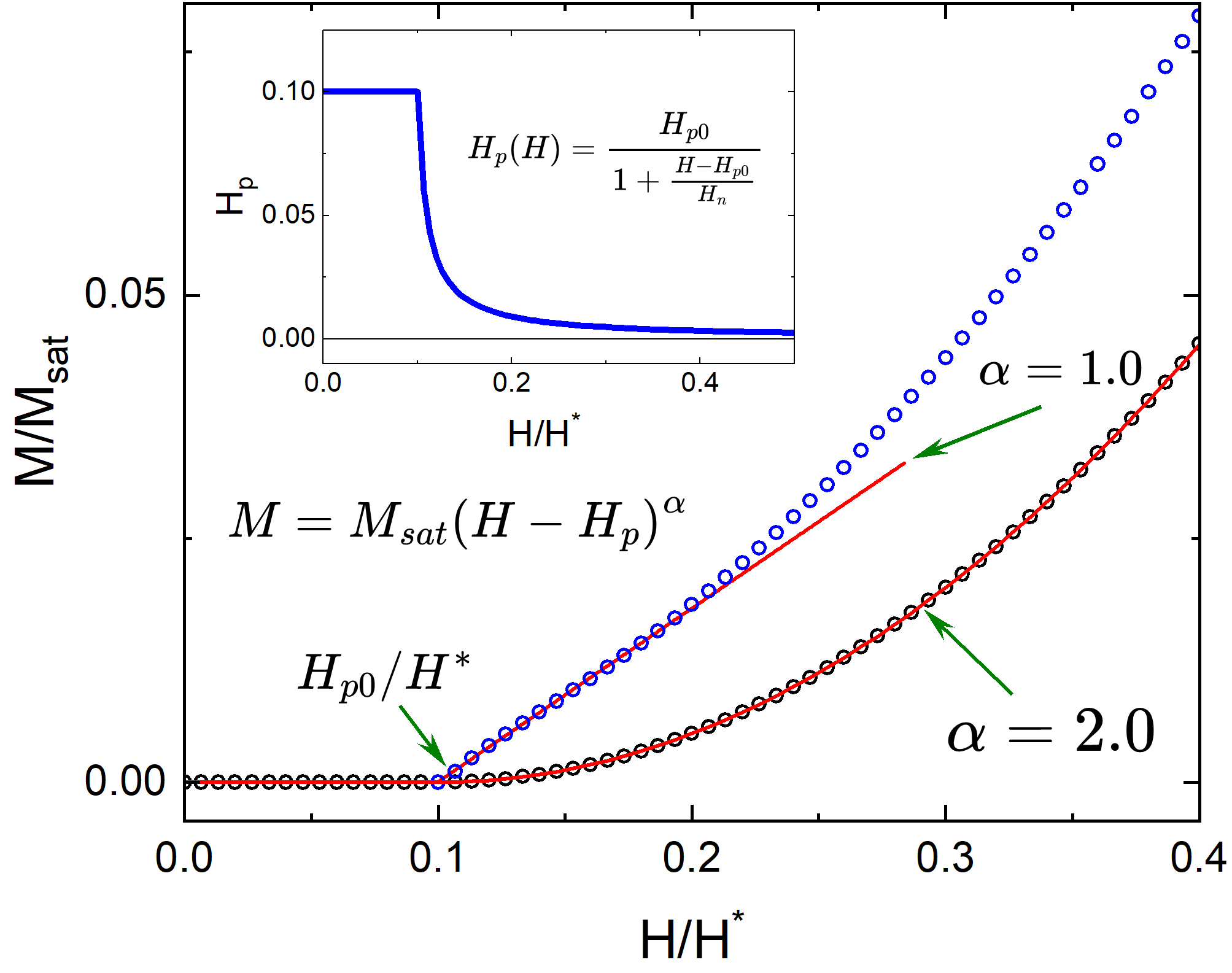}
\caption{Magnetic induction normalized by saturation magnetization. Black circles show Eq.\ref{eq:Mzfc1} with a shifted applied field from Eq.\ref{eq:Hedge}, with a fixed offset, $H_{p}\left(H\right)=H_{p0}=0.1$. Blue circles show Eq.\ref{eq:Mzfc1} with $H_{e}$ obtained from Eq.\ref{eq:Hedge} and $H_{p}\left(H\right)$ from Eq.\ref{eq:Hp} with $H_{p0}=0.1$ and $H_{n}=0.01$. Inset shows $H_{p}\left(H\right)$ for the same parameters.}
\label{fig:Hp(H)}
\end{figure}

\subsubsection{Reversible magnetization}

An important feature of the experimental data is that the trapped magnetic moment starts to deviate from zero only at some finite magnetic field, labeled $H_{p}$ \cite{Budko2023,Minkov2023}. In fact, even in the simplest picture, Abrikosov vortices start to penetrate the bulk at the low critical field $H_{c1}$, or another similar quantity determined, for example, by the surface barrier \cite{Bean1964a}. The total magnetization is the sum of the reversible contribution and the irreversible. The discussion in Section\ref{subsec:Irreversible-magnetization-only} above did not include the reversible part because when a magnetic field is turned off, the reversible component is zero. However, the magnetic field in the ZFC cases should be understood as an effective field shifted by $H_{p}$. When an external magnetic field is applied after cooling in zero field, there is an exponentially attenuated magnetic induction layer with the width given by the London penetration depth, $\lambda$, much smaller than the sample dimensions, so we can approximate it by a ``step'' in magnetic induction on the edge. Although the $H_{p}$ offset was acknowledged in previous work, an important point was overlooked: this edge field step \emph{decreases} with the \emph{increase} of an applied magnetic field. Basically, since the reversible magnetization is $M_{rev}=d^{-1}\int_{0}^{d}(B-H)dV=H_{p}$, the size of the edge step, $H_{p}\left(H\right)$, decreases the same way as the reversible magnetization curve, $M_{rev}(H)$. Therefore, the magnetic field in the previous section for zero-field cooled magnetization should be considered with the magnetic field replaced by
\begin{equation}
H_{e}=H-H_{p}\left(H\right)
\label{eq:Hedge}
\end{equation}
To see how this changes the behavior, let us compare the ZFC-1 trapped magnetization with and without field-dependent edge step. Unfortunately, there is no known analytical form of $M_{rev}(H)$ just above $H_{c1}$. Let us assume that the edge step changes as,
\begin{equation}
H_{p}\left(H\right)=\frac{H_{p0}}{1+\frac{H-H_{p0}}{H_{n}}},\qquad H_{0}\leq H\label{eq:Hp}
\end{equation}
\noindent where $H_{0}$ is a constant of the order of the lower critical field, $H_{c1}$, and $H_{n}$ is some normalization parameter that in real superconductors depends on Ginzburg-Landau parameter $\kappa$. This shape is shown in the insert in Fig.\ref{fig:Hp(H)} with $H_{p0}=0.1$ and $H_{n}=0.01$, which we used later to fit the data. Then, ZFC-1 magnetization is evaluated from Eq,\ref{eq:Mzfc1} with $H_{e}$ obtained from Eq.\ref{eq:Hedge} and Eq.\ref{eq:Hp}. To demonstrate the importance of this correction, Fig.\ref{fig:Hp(H)} compares the unmodified ZFC-1 curve where a constant $H_{p0}=0.1$ was used (black circles), with the curve with $H_{p}(H)$ from Eq.\ref{eq:Hp} (blue circles). The curves are plotted as scatter graphs to show the fits to the power law, $M=M_{sat}\left(H-H_{p0}\right)^{\alpha}$. Naturally, the unmodified curve follows Eq.\ref{eq:Mzfc1} with a fixed exponent $\alpha=2$. However, when the reversible magnetization, shown in the inset, is taken into account the result exhibits close to an $H-$linear behavior at low fields, so that $\alpha\approx1$. Therefore, taking into account reversible magnetization explains the spread of the exponent values, $\alpha=1-2$ observed in known \cite{Budko2023} and UHTS \cite{Minkov2023} superconductors. 

The exact field dependence in Eq.\ref{eq:Hp} is not important. The mere fact that $H_{p}\left(H\right)$ is a \emph{decreasing} function of a magnetic field lowers the exponent $\alpha$ to below $\alpha=2$. In principle, demagnetization correction in platelet samples in the perpendicular field modifies the edge field, and the whole induction profile becomes very different from the slab in a parallel field. Therefore, we expect that the exponent $\alpha$ will be orientation dependent, which is in fact the case \cite{Budko2023}.

We note that even without the edge field, it is not easy to assign any particular exponent when the data are analyzed in the form of $M_{trap}\propto H^{\alpha}$ in the full range of fields. As seen in Fig.\ref{fig:Bean4panels}(d), the combined ZFC-1 plus ZFC-2 curve exhibits a regime close to $H-$ linear in a significant range of intermediate fields. Taking into account severe experimental limitations, noise, realistic sample structure, and, importantly, only a very limited number of data points, the effective extracted exponent $\alpha$ can be significantly lower than 2, even without the edge field correction. 
\footnote{An important note on the number of data points: Each ZFC measurement requires warming above $T_{c}$, making sure that no residual magnetic field is left in a (usually used) superconducting magnet, then cooling back to a low temperature, applying, and turning off the desired magnetic field. With $T_{c}\sim200\:\text{K}$, each point may easily take several hours to collect. The FC points are not much faster, just minus the magnet de-magnetization. Importantly, the sample should remain unchanged during these extended multiple temperature swings, which is not guaranteed for the hydrides that are synthesized inside the pressure cell.}

\subsubsection{Effects of demagnetization}
\label{subsubsec:Effects of demagnetization}

As we have already noted, in discussions of the trapped flux, demagnetizing effects must be important, especially in the case of thin hydride samples. For example, for a disc of $t=3\:\mu\text{m}$ thick and $d=30\:\mu\text{m}$ in diameter, the aspect ratio $\eta=3/30=0.1$ yields for the magnetic field perpendicular to the disc $N_{\perp}=0.83$, which gives the edge enhancement factor $E_{\perp}=1/\left(1-N\right)=5.85$ \cite{Prozorov2018}. The same sample in parallel orientation has $N_{\parallel}=0.06$ and a negligible enhancement, $E_{\parallel}=1/\left(1-N\right)=1.06$. In experiments with H$_{3}$ S the magnetic field was oriented perpendicular to the disc \cite{Minkov2023}. The edge field and magnetic induction penetration profiles in this case are very different from the infinite sample in a parallel field \cite{Brandt1996}. Moreover, all vector components of magnetic induction are important. In fact, in a thin case, the critical state is mostly supported by pinning of the in-plane component, $M_{trap}\propto2B_{r}/t$, where $B_{r}$ is the tangential component of the magnetic inductance on the top and bottom surfaces, and $t$ is the thickness. Furthermore, if a sample cross section (parallel to the field cut in a $xz-$plane) is not ellipsoidal, which is the case for all real samples, magnetic field along the edges in highly non-uniform \cite{Prozorov2021} and flux penetration starts from the corners. There is no longer a well-defined sharp ``first penetration'' field, and solving the incomplete trapping problem becomes difficult. As vortices move in, the cross-sectional shape of the vortex-free volume changes toward a shrinking ellipsoid, and the field dependence of the edge field will be different compared to the infinite case. This means that it is not easy to take the demagnetizing factors into account. The usual renormalization of the edge magnetic field by a factor of $1/(1-N)$ is only true for a perfect diamagnetic (ellipsoidal) sample. The actual correction factor is $1/(1-N\chi)$, where $\chi$ is the magnetic susceptibility of an infinite sample made of the same material. Therefore, we expect the conventional $1/(1-N)$ correction to be applicable in low magnetic fields but to fail in larger fields.

Although the general theory for flat non-ellipsoidal sample is challenging, numerical analysis of this situation, including reversible and irreversible components, finite size and even field-dependent current density is possible along the lines described in great detail by Ernst H. Brandt \cite{Brandt1996}.

\begin{figure}[tb]
\includegraphics[width=8.5cm]{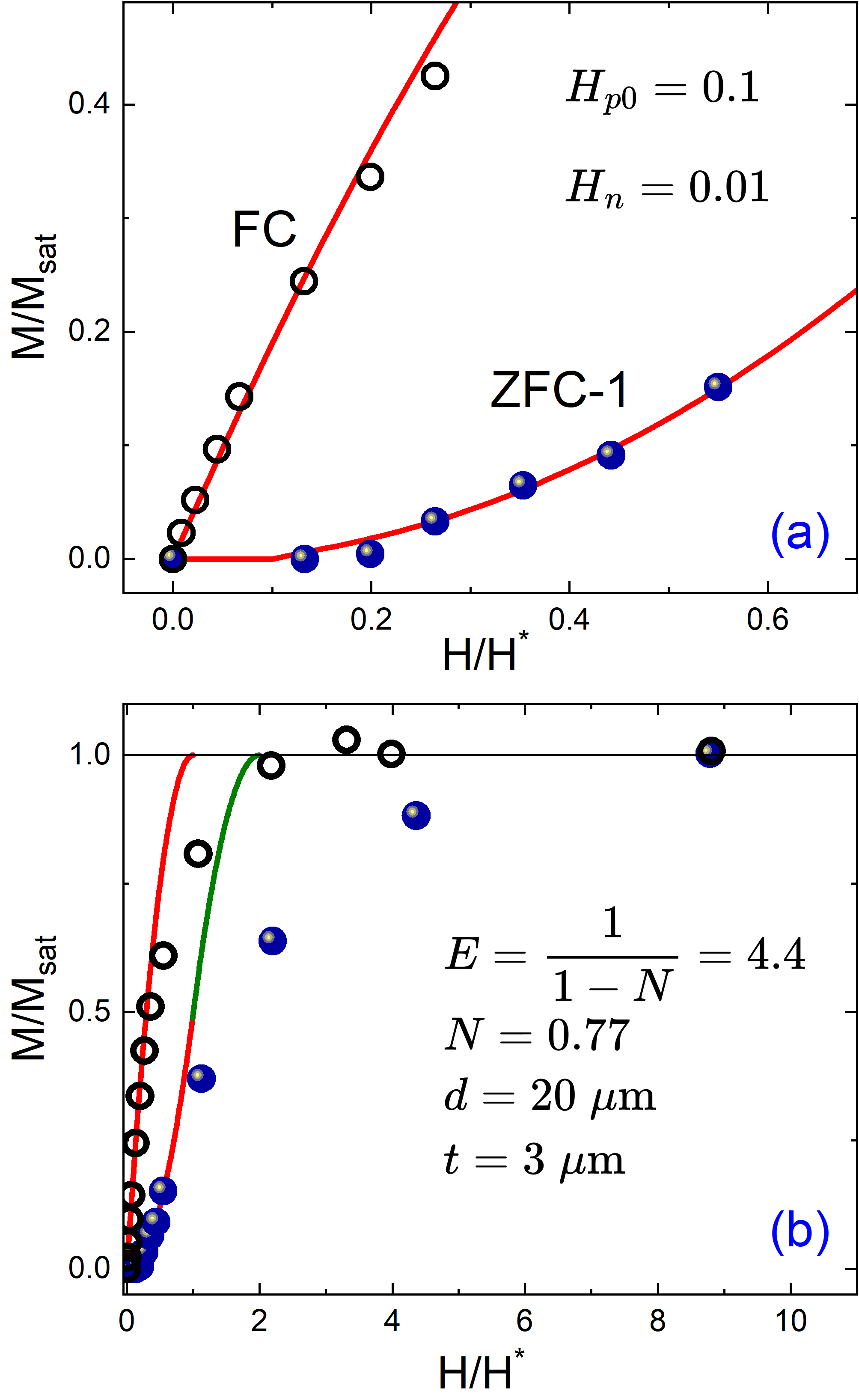}
\caption{Field-dependent magnetization. Open circles show digitized H$_{3}$S data from Fig.2 of Ref.\cite{Minkov2023} with renormalized $H\rightarrow4.4H$ chosen to match our Eq.\ref{eq:Mzfc1}. Open black circles show the FC and filled blue circles show the ZFC data. The lines are from Eq.\ref{eq:BeanFC} for FC, and Eq.\ref{eq:Mzfc1} and Eq.\ref{eq:Mzfc2} with the reversible correction, Eq.\ref{eq:Hedge} and Eq.\ref{eq:Hp}. Numerical values of the parameters used are shown on the graphs. Panel (a) shows the lower fields range and panel (b) shows the expanded view.}
\label{fig:MinkovFit}
\end{figure}

\subsection{Analysis of the trapped magnetic moment experiments in H$_{3}$S from Ref.\cite{Minkov2023}}

In the experiments on H$_{3}$S magnetic field was oriented perpendicular to the disc and the estimates of disc geometry from optical microscopy and X-ray diffraction gave $t=3\:\mu\text{m}$ and $d=85\:\mu\text{m}$ \cite{Minkov2023}. This aspect ratio, $t/d=0.035$, corresponds to a demagnetizing factor, $N=0.93$ and a magnetic field enhancement factor, $E_{\perp}=1/\left(1-N\right)=14.0$ \cite{Prozorov2018}. The further analysis of $H_{c1}$ was more consistent with $E_{\perp}=8.5$, which yields $N=0.88$. For the indicated thickness, the diameter would be $d=48\:\mu\mathrm{m}$. This is quite plausible and the authors list possible reasons \cite{Minkov2023}. In our view, it is likely that the sample is not completely homogeneous in the plane considering its small thickness. It should probably be viewed as a few regions, each behaving separately. All of these estimates are similar by the order of magnitude.

To fit the experimental data, it is important to realize that demagnetization effects enter not only the offset $H_{p}$, but they renormalize the entire $H-$axis. As discussed in Section \ref{subsubsec:Effects of demagnetization}, a constant effective demagnetizing factor is a valid approximation only when the magnetic field does not penetrate too far, so that the aspect ratio of the screened volume does not change much. For larger magnetic fields, the Meissner volume inside shrinks and the enhancement coefficient decreases, meaning that the demagnetizing factor decreases as well. On the other hand, prediction of the trapped moment as a function of field, Eq.\ref{eq:Mzfc1} and Eq.\ref{eq:Mzfc2} are quite rigid and the only flexibility we have is in the shift $H_{p}\left(H\right)$ and a scaling factor $H^\ast$. In order to match our equations, the experimental $H-$axis must be multiplied by the enhancement factor $E_{\perp}$ and divided by $H^\ast$. We had to multiply the $H-$axis of Fig.2 of Ref.\cite{Minkov2023} by a factor 4.4 to achieve the best agreement with our theory, Eq.\ref{eq:Mzfc1}. Note that a single scaling parameter was used for both FC and ZFC data of Fig.2 of Ref.\cite{Minkov2023}. The vertical axis was not changed. The experimental data are plotted together with Eq.\ref{eq:BeanFC} for FC, and Eq.\ref{eq:Mzfc1} and Eq.\ref{eq:Mzfc2} with reversible correction, Eq.\ref{eq:Hedge} and Eq.\ref{eq:Hp}, taken into account in  Figure \ref{fig:MinkovFit}. Panel (a) shows quite a remarkable agreement between theory and experiment for low fields. Note that the general position of the data comes from the demagnetizing effects and the full penetration field, $H^\ast$, while the correct temperature dependence (i.e., exponent $\alpha \approx 1$) comes from the reversible edge step correction. Figure \ref{fig:MinkovFit}(b) shows an expanded view. Expectedly, higher fields do not fit the predicted behavior because of the reasons mentioned in Section \ref{subsubsec:Effects of demagnetization}. 

\subsection{The Kim model}

One must be careful with the quantitative analysis of the exponents based on a simple Bean model, which is practically never observed in real materials. There is always some field dependence of the critical current density, $j_{c}\left(B\right)$. However, it all depends on how significant the field dependence is for the magnetic induction variation across the sample. Usually, these dependencies are weak and while the entire magnetization loop cannot be reproduced by the Bean model, the minor loops with a field span of the order of $2H^{*}$ can still be described by it. In other words, the question is how significant the deviation of the $B\left(x\right)$ profiles from linear is, Fig.\ref{fig:Bean4panels}(a-c). The field dependence is usually most prominent at low fields, which is the regime considered here. Let us review some typical scenarios.

\subsubsection{The original Kim model}

In the Kim model, the current density, $j_{c}$, depends on a magnetic field. The traditional form is
\begin{equation}
j_{c}=\frac{j_{c0}}{1+\beta B}\label{eq:jcBKim}
\end{equation}
\noindent where $j_{c0}$ is the critical current at $B=0$ and the new parameter $\beta$ controls the field dependence contribution. Naturally, the Bean model is obtained from Eq.\ref{eq:jcBKim} by setting $\beta=0$. Let us evaluate the trapped magnetic moment after a partial flux entry, ZFC-1 process. For simplicity, we use the dimensionless variable $x\rightarrow x/d$ . As before, magnetic flux reaches the center at some characteristic magnetic field of full penetration, $H^{*}$, different from the case of the Bean model $H_{Bean}^{*}=dj_{c0}$. Note that we did not include the additional factors associated with the demagnetizing factor and the reversible magnetization via $H_{p}\left(H\right)$, although they will play a similar role as in the case of the Bean model.

\begin{figure}[tb]
\includegraphics[width=8.5cm]{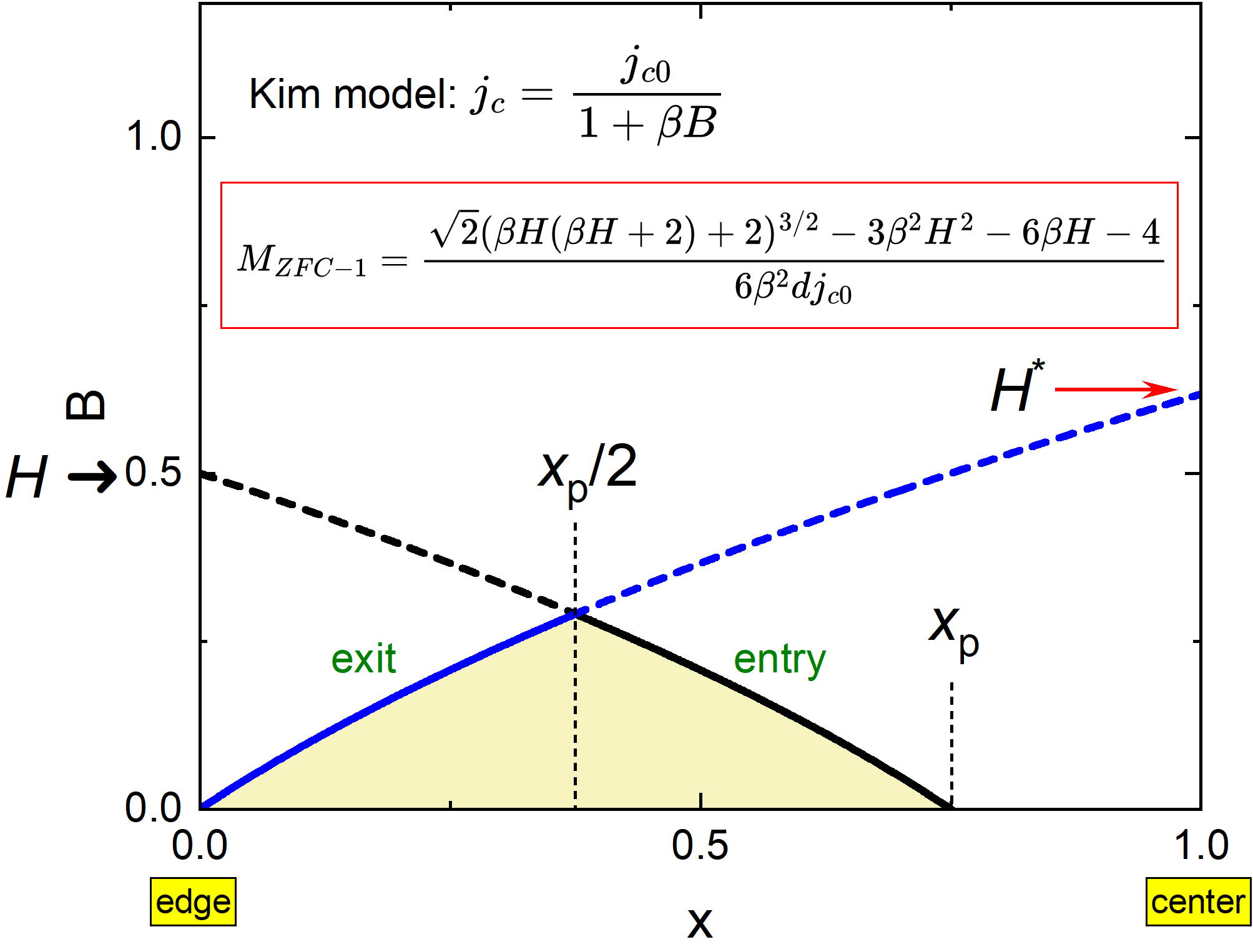} 
\caption{Magnetic induction profiles for the Kim model, Eq.\ref{eq:jcBKim}, in the initial penetration of ZFC-1 process. Parameters for the curves are: $H=0.5$, $\beta=2$ and $j_{c0}=1$.}
\label{fig:KimProfiles}
\end{figure}

First, we need to find the magnetic induction profile inside the sample for a fixed field $H$. Using the Maxwell equation, applicable to our one-dimensional situation, $j=-dB/dx$ and the field-dependent critical current, Eq.\ref{eq:jcBKim} we obtain the differential equation for magnetic induction, $B\left(x\right)$,  
\begin{equation}
\frac{dB}{dx}+\beta\frac{dB}{dx}B+\mu_{0}j_{c0}=0\label{eq:DiffEqKim}
\end{equation}
\noindent with the boundary condition, $B\left(x=0\right)=H$. The solution of this equation is straightforward, 
\begin{equation}
B\left(x\right)=\frac{1}{\beta}\frac{\sqrt{1+\beta^{2}H^{2}+2\beta H-2\beta j_{c0}x}}{\alpha}-1\label{eq:B(x)Kim}
\end{equation}
The Bean model is obtained from Eq.\ref{eq:B(x)Kim} by evaluating
\begin{equation}
B_{Bean}\left(x\right)=\lim_{\beta\rightarrow0}B\left(x\right)=H-j_{c0}x\label{eq:B(x)KimBean}
\end{equation}
\noindent a straight line, as expected, see Fig.\ref{fig:Bean4panels}(b). Figure \ref{fig:KimProfiles} shows $B\left(x\right)$ profiles corresponding to Eq.\ref{eq:jcBKim} for some combination of parameters, $H=0.5$, $\beta=2$ and $j_{c0}=1$. The flux entry profile is labeled ``entry'' and the flux exit profile, labeled ``exit'', is obtained from Eq.\ref{eq:B(x)Kim} by inverting the $x-$coordinate, $x'\rightarrow x_{p}-x$. The shaded area, $A$, is proportional to the trapped magnetic moment. The penetration distance, $x_{p}$, is found by solving $B\left(x_{p}\right)=0$, Eq.\ref{eq:B(x)Kim},
\begin{equation}
x_{p}=\frac{\beta H^{2}+2H}{2j_{c0}}\label{eq:xpKim}
\end{equation}

The field at which flux reaches the center, the field of full penetration, $H^{*}$, is determined from $x_{p}=1$, 
\begin{equation}
H^{*}=\frac{1}{\beta}\left(\sqrt{1+2j_{c0}\beta}-1\right)\label{eq:HpKim}
\end{equation}

In the limit of field-independent current density we obtain $H_{Bean}^{*}=j_{c0}$, as expected (this field is $j_{c0}x_{p}$, where $x_{p}=d=1$ in this case, but in real units it is given by the usual $H_{Bean}^{*}=dj_{c0}$). The magnetic field $H^{*}$ is a maximum field in which this ZFC-1 model is applicable. Above it, the magnetic moment should be calculated differently, because the profiles are no longer symmetric with respect to $x_{p}/2$. Substituting Eq.\ref{eq:B(x)Kim} into Eq.\ref{eq:Mtrap}, we obtain,
\begin{equation}
M_{ZFC-1}=\frac{\sqrt{2}(\beta H(\beta H+2)+2)^{3/2}-3\beta^{2}H^{2}-6\beta H-4}{6\beta^{2}dj_{c0}}\label{eq:KimZFC-1}
\end{equation}

Checking the Bean model limit we have,  
\begin{equation}
M_{t,Bean}=\lim_{\beta\rightarrow0}M_{t}=\frac{H^{2}}{4dj_{c0}}=\frac{H^{2}}{4H_{Bean}^{*}}\label{eq:MKimBeanLimit}
\end{equation}
\noindent as expected from the constant current density model, Eq.\ref{eq:Mzfc1}.

To compare with the experimentally derived power-law dependence, $M_{ZFC-1}\propto H^{\alpha}$, we expand Eq.\ref{eq:KimZFC-1} in the powers of $H$,
\begin{equation}
M_{approx}\approx\frac{H^{2}}{4H_{Bean}^{*}}\left(1+\frac{5}{6}\beta H\right)+O\left(H^{4}\right)\label{eq:MtrapKimPowerLaw}
\end{equation}
(We recall that, according to Eq.\ref{eq:jcBKim}, $\beta H$ is dimensionless).
Therefore, in the classical Kim model, we expect the exponent $\alpha$
to lie in the interval $2\leq\alpha<3$, so, it differs from the pure $H^{2}$ law predicted by the Bean model, Eq.\ref{eq:Mzfc1}, with a field-independent current density. Apparently, decreasing with field critical current expands the upper limit of possible exponents. An important general conclusion is that this exponent is not fixed at $\alpha=2$, but depends on the parameters of the model. To further explore this conclusion, let us consider a modified Kim model.

\subsubsection{The modified Kim Model}

The field dependence of the critical current density can be different from Eq.\ref{eq:jcBKim}. To examine how sensitive this is, let us consider a modified Kim model,
\begin{equation}
j_{c}=\frac{j_{c0}}{1+\beta B^{2}}\label{eq:jcBKim2}
\end{equation}
\noindent which differs from Eq.\ref{eq:jcBKim} by the quadratic $B$ term in the denominator. Following the procedure similar to previous section, the magnetic induction profile is found from the solution of the following differential equation, 
\begin{equation}
\frac{dB}{dx}+\beta\frac{dB}{dx}B^{2}+\mu_{0}j_{c0}=0\label{eq:DiffEqKim2}
\end{equation}
\noindent with the boundary condition, $B\left(x=0\right)=H$. The analytical solution of Eq.\ref{eq:DiffEqKim2} in Mathematica is straightforward, but the formula is too cumbersome to reproduce here. It is easy to check that it does produce a correct linear profile in the Bean $\beta\rightarrow0$ limit. The flux penetration distance, $x_{p}$ from the edge, is found
from $B\left(x_{p}\right)=0$, yielding: 
\begin{equation}
x_{p}=\frac{H\left(\beta H^{2}+3\right)}{3j_{c0}}\label{eq:xpKim-1}
\end{equation}

The trapped magnetic moment is given by Eq.\ref{eq:Mtrap} with $B\left(x\right)$ calculated as a solution of Eq.\ref{eq:DiffEqKim2}. Again, the final analytical expression is too bulky to write here, but it can be readily obtained in Mathematica following the described procedure. The power law expansion, however, is short enough,
\begin{equation}
M_{ZFC-1}\approx\frac{H^{2}}{4H_{Bean}^{*}}\left(1+\frac{5}{8}\beta H^{2}\right)+O\left(H^{5}\right)\label{eq:MtrapKim2PowerLaw}
\end{equation}
(Here $\beta H^{2}$ is dimensionless, see Eq.\ref{eq:jcBKim2}). Therefore, in the modified Kim model, Eq.\ref{eq:jcBKim2}, we expect the exponent $\alpha$ in the effective power-law behavior of the trapped magnetic flux after zero-field cooling, $M_{t,a}\propto H^{\alpha}$ to lie even in the larger interval, $2\leq\alpha<4$. It differs from the rigid $H^{2}$ behavior, predicted by the Bean model with constant current density, and further expands the range of exponents possible in this experiment.

\section{Conclusions}

In conclusion, we analyzed possible outcomes of the flux trapping experiments in which a small magnetic field, $H$, is turned off after field cooling (FC), or zero-field cooling (ZFC-1 and ZFC-2 protocols) and the trapped magnetic moment is measured. Analytical expressions are derived for the classical Bean ($j_{c}\neq f(B)$) and Kim ($j_{c}\propto1/(1+\beta B^{(1,2)})$) models and analyzed in terms of the power-law behavior of trapped magnetization, $M_{trap}\propto H^{\alpha}$, which is a convenient function to approximate the experimental results. While for the Bean model, the exponent is fixed, $\alpha=2$, the Kim models expand the range of possible exponents in a wide interval $2\leq\alpha\leq4$. With reversible magnetization taken into account, the effective exponent can be as low as $\alpha=1$ (possibly lower). Furthermore, the demagnetizing factor corrections introduce an orientation dependence of the results, in addition to possible crystalline anisotropies. Together, these factors explain the spread of the experimental values of $\alpha$ obtained in known superconductors \cite{Budko2023} and H$_3$S UHTS \cite{Minkov2023}. 

Since this work was largely motivated by flux trapping experiments on H$_{3}$S under ultra-high pressure \cite{Minkov2022}, their data were digitized and analyzed within the presented theory. It is shown that the derived expressions fit the data rather well, providing further confirmation of the bulk type-II superconductivity in this compound. The results of this work are in line with our earlier conclusions of the applicability of the critical state model to H$_{3}$S, derived from a completely different experiment of flux shielding studied by M\"{o}ssbauer spectroscopy \cite{Eremets2022,Prozorov2022}.

Finally, as a side note, curves similar to Fig.\ref{fig:Bean4panels}(d) can be obtained if similar measurements were performed on a ferromagnetic sample with pinning. In this case, the remanent magnetic moment is also positive and its field dependence is due to the $H-$dependent size of the minor $M\left(H\right)$ loops when an applied magnetic field is insufficient to fully remagnetize the sample (i.e. reorient all magnetic domains). In this case, the demonstration of a diamagnetic response is imperative. In case of H$_{3}$S, diamagnetic screening was demonstrated before \cite{Minkov2022} the flux trapping experiments \cite{Minkov2023}. 

\begin{acknowledgments}
I thank S. Bud'ko, P. Canfield, V. Kogan, and J. Hirsch for discussions. This work was supported by the U.S. Department of Energy (DOE), Office of Science, Basic Energy Sciences, Division of Materials Science and Engineering. Ames Laboratory is operated for the U.S. DOE by Iowa State University under the contract DE-AC02-07CH11358. 
\end{acknowledgments}
%
%\bibliography{KimModel}

\begin{thebibliography}{24}%
\makeatletter
\providecommand \@ifxundefined [1]{%
 \@ifx{#1\undefined}
}%
\providecommand \@ifnum [1]{%
 \ifnum #1\expandafter \@firstoftwo
 \else \expandafter \@secondoftwo
 \fi
}%
\providecommand \@ifx [1]{%
 \ifx #1\expandafter \@firstoftwo
 \else \expandafter \@secondoftwo
 \fi
}%
\providecommand \natexlab [1]{#1}%
\providecommand \enquote  [1]{``#1''}%
\providecommand \bibnamefont  [1]{#1}%
\providecommand \bibfnamefont [1]{#1}%
\providecommand \citenamefont [1]{#1}%
\providecommand \href@noop [0]{\@secondoftwo}%
\providecommand \href [0]{\begingroup \@sanitize@url \@href}%
\providecommand \@href[1]{\@@startlink{#1}\@@href}%
\providecommand \@@href[1]{\endgroup#1\@@endlink}%
\providecommand \@sanitize@url [0]{\catcode `\\12\catcode `\$12\catcode
  `\&12\catcode `\#12\catcode `\^12\catcode `\_12\catcode `\%12\relax}%
\providecommand \@@startlink[1]{}%
\providecommand \@@endlink[0]{}%
\providecommand \url  [0]{\begingroup\@sanitize@url \@url }%
\providecommand \@url [1]{\endgroup\@href {#1}{\urlprefix }}%
\providecommand \urlprefix  [0]{URL }%
\providecommand \Eprint [0]{\href }%
\providecommand \doibase [0]{https://doi.org/}%
\providecommand \selectlanguage [0]{\@gobble}%
\providecommand \bibinfo  [0]{\@secondoftwo}%
\providecommand \bibfield  [0]{\@secondoftwo}%
\providecommand \translation [1]{[#1]}%
\providecommand \BibitemOpen [0]{}%
\providecommand \bibitemStop [0]{}%
\providecommand \bibitemNoStop [0]{.\EOS\space}%
\providecommand \EOS [0]{\spacefactor3000\relax}%
\providecommand \BibitemShut  [1]{\csname bibitem#1\endcsname}%
\let\auto@bib@innerbib\@empty
%</preamble>
\bibitem [{\citenamefont {Pickard}\ \emph {et~al.}(2020)\citenamefont
  {Pickard}, \citenamefont {Errea},\ and\ \citenamefont
  {Eremets}}]{Pickard2020}%
  \BibitemOpen
  \bibfield  {author} {\bibinfo {author} {\bibfnamefont {C.~J.}\ \bibnamefont
  {Pickard}}, \bibinfo {author} {\bibfnamefont {I.}~\bibnamefont {Errea}},\
  and\ \bibinfo {author} {\bibfnamefont {M.~I.}\ \bibnamefont {Eremets}},\
  }\bibfield  {title} {\bibinfo {title} {Superconducting hydrides under
  pressure},\ }\href
  {https://doi.org/https://doi.org/10.1146/annurev-conmatphys-031218-013413}
  {\bibfield  {journal} {\bibinfo  {journal} {Annu. Rev. Conden. Ma. P.}\
  }\textbf {\bibinfo {volume} {11}},\ \bibinfo {pages} {57} (\bibinfo {year}
  {2020})}\BibitemShut {NoStop}%
\bibitem [{\citenamefont {Pickett}(2023)}]{Pickett2023}%
  \BibitemOpen
  \bibfield  {author} {\bibinfo {author} {\bibfnamefont {W.~E.}\ \bibnamefont
  {Pickett}},\ }\bibfield  {title} {\bibinfo {title} {Colloquium: Room
  temperature superconductivity: The roles of theory and materials design},\
  }\href {https://doi.org/10.1103/RevModPhys.95.021001} {\bibfield  {journal}
  {\bibinfo  {journal} {Rev. Mod. Phys.}\ }\textbf {\bibinfo {volume} {95}},\
  \bibinfo {pages} {021001} (\bibinfo {year} {2023})}\BibitemShut {NoStop}%
\bibitem [{\citenamefont {Zhao}\ \emph {et~al.}(2023)\citenamefont {Zhao},
  \citenamefont {Huang}, \citenamefont {Zhang}, \citenamefont {Chen},
  \citenamefont {Du}, \citenamefont {Duan},\ and\ \citenamefont
  {Cui}}]{Zhao2023}%
  \BibitemOpen
  \bibfield  {author} {\bibinfo {author} {\bibfnamefont {W.}~\bibnamefont
  {Zhao}}, \bibinfo {author} {\bibfnamefont {X.}~\bibnamefont {Huang}},
  \bibinfo {author} {\bibfnamefont {Z.}~\bibnamefont {Zhang}}, \bibinfo
  {author} {\bibfnamefont {S.}~\bibnamefont {Chen}}, \bibinfo {author}
  {\bibfnamefont {M.}~\bibnamefont {Du}}, \bibinfo {author} {\bibfnamefont
  {D.}~\bibnamefont {Duan}},\ and\ \bibinfo {author} {\bibfnamefont
  {T.}~\bibnamefont {Cui}},\ }\bibfield  {title} {\bibinfo {title}
  {Superconducting ternary hydrides: progress and challenges},\ }\href
  {https://doi.org/10.1093/nsr/nwad307} {\bibfield  {journal} {\bibinfo
  {journal} {Natl. Sci. Rev.}\ }\textbf {\bibinfo {volume} {11}},\ \bibinfo
  {pages} {nwad307} (\bibinfo {year} {2023})}\BibitemShut {NoStop}%
\bibitem [{\citenamefont {Troyan}\ \emph {et~al.}(2024)\citenamefont {Troyan},
  \citenamefont {Semenok}, \citenamefont {Sadakov}, \citenamefont {Lyubutin},\
  and\ \citenamefont {Pudalov}}]{Troyan2024}%
  \BibitemOpen
  \bibfield  {author} {\bibinfo {author} {\bibfnamefont {I.~A.}\ \bibnamefont
  {Troyan}}, \bibinfo {author} {\bibfnamefont {D.~V.}\ \bibnamefont {Semenok}},
  \bibinfo {author} {\bibfnamefont {A.~V.}\ \bibnamefont {Sadakov}}, \bibinfo
  {author} {\bibfnamefont {I.~S.}\ \bibnamefont {Lyubutin}},\ and\ \bibinfo
  {author} {\bibfnamefont {V.~M.}\ \bibnamefont {Pudalov}},\ }\bibfield
  {title} {\bibinfo {title} {Progress, problems and prospects of
  room-temperature superconductivity},\ }\href@noop {} {\bibfield  {journal}
  {\bibinfo  {journal} {arXiv:2406.11344}\ } (\bibinfo {year}
  {2024})}\BibitemShut {NoStop}%
\bibitem [{\citenamefont {Eremets}\ \emph {et~al.}(2022)\citenamefont
  {Eremets}, \citenamefont {Minkov}, \citenamefont {Drozdov}, \citenamefont
  {Kong}, \citenamefont {Ksenofontov}, \citenamefont {Shylin}, \citenamefont
  {Bud’ko}, \citenamefont {Prozorov}, \citenamefont {Balakirev},
  \citenamefont {Sun}, \citenamefont {Mozaffari},\ and\ \citenamefont
  {Balicas}}]{Eremets2022}%
  \BibitemOpen
  \bibfield  {author} {\bibinfo {author} {\bibfnamefont {M.~I.}\ \bibnamefont
  {Eremets}}, \bibinfo {author} {\bibfnamefont {V.~S.}\ \bibnamefont {Minkov}},
  \bibinfo {author} {\bibfnamefont {A.~P.}\ \bibnamefont {Drozdov}}, \bibinfo
  {author} {\bibfnamefont {P.~P.}\ \bibnamefont {Kong}}, \bibinfo {author}
  {\bibfnamefont {V.}~\bibnamefont {Ksenofontov}}, \bibinfo {author}
  {\bibfnamefont {S.~I.}\ \bibnamefont {Shylin}}, \bibinfo {author}
  {\bibfnamefont {S.~L.}\ \bibnamefont {Bud’ko}}, \bibinfo {author}
  {\bibfnamefont {R.}~\bibnamefont {Prozorov}}, \bibinfo {author}
  {\bibfnamefont {F.~F.}\ \bibnamefont {Balakirev}}, \bibinfo {author}
  {\bibfnamefont {D.}~\bibnamefont {Sun}}, \bibinfo {author} {\bibfnamefont
  {S.}~\bibnamefont {Mozaffari}},\ and\ \bibinfo {author} {\bibfnamefont
  {L.}~\bibnamefont {Balicas}},\ }\bibfield  {title} {\bibinfo {title}
  {High-temperature superconductivity in hydrides: Experimental evidence and
  details},\ }\href {https://doi.org/10.1007/s10948-022-06148-1} {\bibfield
  {journal} {\bibinfo  {journal} {J. Supercond. Nov. Magn.}\ }\textbf {\bibinfo
  {volume} {35}},\ \bibinfo {pages} {965} (\bibinfo {year} {2022})}\BibitemShut
  {NoStop}%
\bibitem [{\citenamefont {Minkov}\ \emph {et~al.}(2022)\citenamefont {Minkov},
  \citenamefont {Bud’ko}, \citenamefont {Balakirev}, \citenamefont
  {Prakapenka}, \citenamefont {Chariton}, \citenamefont {Husband},
  \citenamefont {Liermann},\ and\ \citenamefont {Eremets}}]{Minkov2022}%
  \BibitemOpen
  \bibfield  {author} {\bibinfo {author} {\bibfnamefont {V.~S.}\ \bibnamefont
  {Minkov}}, \bibinfo {author} {\bibfnamefont {S.~L.}\ \bibnamefont
  {Bud’ko}}, \bibinfo {author} {\bibfnamefont {F.~F.}\ \bibnamefont
  {Balakirev}}, \bibinfo {author} {\bibfnamefont {V.~B.}\ \bibnamefont
  {Prakapenka}}, \bibinfo {author} {\bibfnamefont {S.}~\bibnamefont
  {Chariton}}, \bibinfo {author} {\bibfnamefont {R.~J.}\ \bibnamefont
  {Husband}}, \bibinfo {author} {\bibfnamefont {H.~P.}\ \bibnamefont
  {Liermann}},\ and\ \bibinfo {author} {\bibfnamefont {M.~I.}\ \bibnamefont
  {Eremets}},\ }\bibfield  {title} {\bibinfo {title} {Magnetic field screening
  in hydrogen-rich high-temperature superconductors},\ }\href
  {https://doi.org/10.1038/s41467-022-30782-x} {\bibfield  {journal} {\bibinfo
  {journal} {Nat. Commun.}\ }\textbf {\bibinfo {volume} {13}},\ \bibinfo
  {pages} {3194} (\bibinfo {year} {2022})}\BibitemShut {NoStop}%
\bibitem [{\citenamefont {Prozorov}\ and\ \citenamefont
  {Bud'ko}(2022)}]{Prozorov2022}%
  \BibitemOpen
  \bibfield  {author} {\bibinfo {author} {\bibfnamefont {R.}~\bibnamefont
  {Prozorov}}\ and\ \bibinfo {author} {\bibfnamefont {S.~L.}\ \bibnamefont
  {Bud'ko}},\ }\bibfield  {title} {\bibinfo {title} {{On the Analysis of the
  Tin-Inside-H$_3$S M{\"{o}}ssbauer Experiment}},\ }\href
  {https://doi.org/10.1007/s10948-022-06371-w} {\bibfield  {journal} {\bibinfo
  {journal} {J. Supercond. Nov. Magn.}\ }\textbf {\bibinfo {volume} {35}},\
  \bibinfo {pages} {2615–2619} (\bibinfo {year} {2022})}\BibitemShut
  {NoStop}%
\bibitem [{\citenamefont {Minkov}\ \emph {et~al.}(2023)\citenamefont {Minkov},
  \citenamefont {Ksenofontov}, \citenamefont {Bud’ko}, \citenamefont
  {Talantsev},\ and\ \citenamefont {Eremets}}]{Minkov2023}%
  \BibitemOpen
  \bibfield  {author} {\bibinfo {author} {\bibfnamefont {V.~S.}\ \bibnamefont
  {Minkov}}, \bibinfo {author} {\bibfnamefont {V.}~\bibnamefont {Ksenofontov}},
  \bibinfo {author} {\bibfnamefont {S.~L.}\ \bibnamefont {Bud’ko}}, \bibinfo
  {author} {\bibfnamefont {E.~F.}\ \bibnamefont {Talantsev}},\ and\ \bibinfo
  {author} {\bibfnamefont {M.~I.}\ \bibnamefont {Eremets}},\ }\bibfield
  {title} {\bibinfo {title} {Magnetic flux trapping in hydrogen-rich
  high-temperature superconductors},\ }\href
  {https://doi.org/10.1038/s41567-023-02089-1} {\bibfield  {journal} {\bibinfo
  {journal} {Nat. Phys.}\ }\textbf {\bibinfo {volume} {19}},\ \bibinfo {pages}
  {1293} (\bibinfo {year} {2023})}\BibitemShut {NoStop}%
\bibitem [{\citenamefont {Hirsch}\ and\ \citenamefont
  {Marsiglio}(2022)}]{Hirsch2022}%
  \BibitemOpen
  \bibfield  {author} {\bibinfo {author} {\bibfnamefont {J.~E.}\ \bibnamefont
  {Hirsch}}\ and\ \bibinfo {author} {\bibfnamefont {F.}~\bibnamefont
  {Marsiglio}},\ }\bibfield  {title} {\bibinfo {title} {Evidence against
  superconductivity in flux trapping experiments on hydrides under high
  pressure},\ }\href {https://doi.org/10.1007/s10948-022-06365-8} {\bibfield
  {journal} {\bibinfo  {journal} {J. Supercond. Nov. Magn.}\ }\textbf {\bibinfo
  {volume} {35}},\ \bibinfo {pages} {3141} (\bibinfo {year}
  {2022})}\BibitemShut {NoStop}%
\bibitem [{\citenamefont {Bud’ko}(2024)}]{Budko2024}%
  \BibitemOpen
  \bibfield  {author} {\bibinfo {author} {\bibfnamefont {S.~L.}\ \bibnamefont
  {Bud’ko}},\ }\bibfield  {title} {\bibinfo {title} {{Comment on "Evidence
  Against Superconductivity in Flux Trapping Experiments on Hydrides Under High
  Pressure"}},\ }\href {https://doi.org/10.1007/s10948-024-06712-x} {\bibfield
  {journal} {\bibinfo  {journal} {J. Supercond. Nov. Magn.}\ }\textbf {\bibinfo
  {volume} {37}},\ \bibinfo {pages} {473} (\bibinfo {year} {2024})}\BibitemShut
  {NoStop}%
\bibitem [{\citenamefont {Bud’ko}\ \emph {et~al.}(2024)\citenamefont
  {Bud’ko}, \citenamefont {Huyan}, \citenamefont {Xu},\ and\ \citenamefont
  {Canfield}}]{Budko2024a}%
  \BibitemOpen
  \bibfield  {author} {\bibinfo {author} {\bibfnamefont {S.~L.}\ \bibnamefont
  {Bud’ko}}, \bibinfo {author} {\bibfnamefont {S.}~\bibnamefont {Huyan}},
  \bibinfo {author} {\bibfnamefont {M.}~\bibnamefont {Xu}},\ and\ \bibinfo
  {author} {\bibfnamefont {P.~C.}\ \bibnamefont {Canfield}},\ }\bibfield
  {title} {\bibinfo {title} {{Trapped flux in a small crystal of
  CaKFe$_4$As$_4$ at ambient pressure and in a diamond anvil pressure cell}},\
  }\href {https://doi.org/10.1088/1361-6668/ad45c7} {\bibfield  {journal}
  {\bibinfo  {journal} {Superconductor Science and Technology}\ }\textbf
  {\bibinfo {volume} {37}},\ \bibinfo {pages} {065010} (\bibinfo {year}
  {2024})}\BibitemShut {NoStop}%
\bibitem [{\citenamefont {Bud’ko}\ \emph {et~al.}(2023)\citenamefont
  {Bud’ko}, \citenamefont {Xu},\ and\ \citenamefont {Canfield}}]{Budko2023}%
  \BibitemOpen
  \bibfield  {author} {\bibinfo {author} {\bibfnamefont {S.~L.}\ \bibnamefont
  {Bud’ko}}, \bibinfo {author} {\bibfnamefont {M.}~\bibnamefont {Xu}},\ and\
  \bibinfo {author} {\bibfnamefont {P.~C.}\ \bibnamefont {Canfield}},\
  }\bibfield  {title} {\bibinfo {title} {{Trapped flux in pure and
  Mn-substituted CaKFe$_4$As$_4$ and MgB$_2$ superconducting single
  crystals}},\ }\href {https://doi.org/10.1088/1361-6668/acf413} {\bibfield
  {journal} {\bibinfo  {journal} {Supercond. Sci. Technol.}\ }\textbf {\bibinfo
  {volume} {36}},\ \bibinfo {pages} {115001} (\bibinfo {year}
  {2023})}\BibitemShut {NoStop}%
\bibitem [{\citenamefont {Hirsch}\ and\ \citenamefont
  {Marsiglio}(2024{\natexlab{a}})}]{Hirsch2024}%
  \BibitemOpen
  \bibfield  {author} {\bibinfo {author} {\bibfnamefont {J.~E.}\ \bibnamefont
  {Hirsch}}\ and\ \bibinfo {author} {\bibfnamefont {F.}~\bibnamefont
  {Marsiglio}},\ }\bibfield  {title} {\bibinfo {title} {{Comment on "Trapped
  flux in a small crystal of CaKFe$_4$As$_4$ at ambient pressure and in a
  diamond anvil pressure cell" by S. L. Bud'ko et al}},\ }\href@noop {}
  {\bibfield  {journal} {\bibinfo  {journal} {arXiv:2405.17500}\ } (\bibinfo
  {year} {2024}{\natexlab{a}})}\BibitemShut {NoStop}%
\bibitem [{\citenamefont {Hirsch}\ and\ \citenamefont
  {Marsiglio}(2024{\natexlab{b}})}]{Hirsch2024a}%
  \BibitemOpen
  \bibfield  {author} {\bibinfo {author} {\bibfnamefont {J.~E.}\ \bibnamefont
  {Hirsch}}\ and\ \bibinfo {author} {\bibfnamefont {F.}~\bibnamefont
  {Marsiglio}},\ }\bibfield  {title} {\bibinfo {title} {Further analysis of
  flux trapping experiments on hydrides under high pressure},\ }\href
  {https://doi.org/https://doi.org/10.1016/j.physc.2024.1354500} {\bibfield
  {journal} {\bibinfo  {journal} {Physica C}\ }\textbf {\bibinfo {volume}
  {620}},\ \bibinfo {pages} {1354500} (\bibinfo {year}
  {2024}{\natexlab{b}})}\BibitemShut {NoStop}%
\bibitem [{\citenamefont {Bean}(1962)}]{Bean1962}%
  \BibitemOpen
  \bibfield  {author} {\bibinfo {author} {\bibfnamefont {C.~P.}\ \bibnamefont
  {Bean}},\ }\bibfield  {title} {\bibinfo {title} {{M}agnetization of {H}ard
  {S}uperconductors},\ }\href {http://link.aps.org/abstract/PRL/v8/p250}
  {\bibfield  {journal} {\bibinfo  {journal} {Phys. Rev. Lett.}\ }\textbf
  {\bibinfo {volume} {8}},\ \bibinfo {pages} {250} (\bibinfo {year}
  {1962})}\BibitemShut {NoStop}%
\bibitem [{\citenamefont {Bean}(1964)}]{Bean1964}%
  \BibitemOpen
  \bibfield  {author} {\bibinfo {author} {\bibfnamefont {C.~P.}\ \bibnamefont
  {Bean}},\ }\bibfield  {title} {\bibinfo {title} {{M}agnetization of
  {H}igh-{F}ield {S}uperconductors},\ }\href
  {http://link.aps.org/abstract/RMP/v36/p31} {\bibfield  {journal} {\bibinfo
  {journal} {Rev. Mod. Phys.}\ }\textbf {\bibinfo {volume} {36}},\ \bibinfo
  {pages} {31} (\bibinfo {year} {1964})}\BibitemShut {NoStop}%
\bibitem [{\citenamefont {London}(1963)}]{London1963}%
  \BibitemOpen
  \bibfield  {author} {\bibinfo {author} {\bibfnamefont {H.}~\bibnamefont
  {London}},\ }\bibfield  {title} {\bibinfo {title} {Alternating current losses
  in superconductors of the second kind},\ }\href
  {https://doi.org/10.1016/0031-9163(63)90527-4} {\bibfield  {journal}
  {\bibinfo  {journal} {Phys. Lett.}\ }\textbf {\bibinfo {volume} {6}},\
  \bibinfo {pages} {162} (\bibinfo {year} {1963})}\BibitemShut {NoStop}%
\bibitem [{\citenamefont {Kim}\ \emph {et~al.}(1962)\citenamefont {Kim},
  \citenamefont {Hempstead},\ and\ \citenamefont {Strnad}}]{Kim1962}%
  \BibitemOpen
  \bibfield  {author} {\bibinfo {author} {\bibfnamefont {Y.~B.}\ \bibnamefont
  {Kim}}, \bibinfo {author} {\bibfnamefont {C.~F.}\ \bibnamefont {Hempstead}},\
  and\ \bibinfo {author} {\bibfnamefont {A.~R.}\ \bibnamefont {Strnad}},\
  }\bibfield  {title} {\bibinfo {title} {Critical persistent currents in hard
  superconductors},\ }\href {https://doi.org/10.1103/physrevlett.9.306}
  {\bibfield  {journal} {\bibinfo  {journal} {Phys. Rev. Lett.}\ }\textbf
  {\bibinfo {volume} {9}},\ \bibinfo {pages} {306} (\bibinfo {year}
  {1962})}\BibitemShut {NoStop}%
\bibitem [{\citenamefont {Kim}\ \emph {et~al.}(1963)\citenamefont {Kim},
  \citenamefont {Hempstead},\ and\ \citenamefont {Strnad}}]{Kim1963}%
  \BibitemOpen
  \bibfield  {author} {\bibinfo {author} {\bibfnamefont {Y.~B.}\ \bibnamefont
  {Kim}}, \bibinfo {author} {\bibfnamefont {C.~F.}\ \bibnamefont {Hempstead}},\
  and\ \bibinfo {author} {\bibfnamefont {A.~R.}\ \bibnamefont {Strnad}},\
  }\bibfield  {title} {\bibinfo {title} {Magnetization and critical
  supercurrents},\ }\href {https://doi.org/10.1103/physrev.129.528} {\bibfield
  {journal} {\bibinfo  {journal} {Phys. Rev.}\ }\textbf {\bibinfo {volume}
  {129}},\ \bibinfo {pages} {528} (\bibinfo {year} {1963})}\BibitemShut
  {NoStop}%
\bibitem [{\citenamefont {Chen}\ and\ \citenamefont
  {Goldfarb}(1989)}]{Chen1989}%
  \BibitemOpen
  \bibfield  {author} {\bibinfo {author} {\bibfnamefont {D.-X.}\ \bibnamefont
  {Chen}}\ and\ \bibinfo {author} {\bibfnamefont {R.~B.}\ \bibnamefont
  {Goldfarb}},\ }\bibfield  {title} {\bibinfo {title} {Kim model for
  magnetization of type-ii superconductors},\ }\href
  {https://doi.org/10.1063/1.344261} {\bibfield  {journal} {\bibinfo  {journal}
  {J. Appl. Phys.}\ }\textbf {\bibinfo {volume} {66}},\ \bibinfo {pages} {2489}
  (\bibinfo {year} {1989})}\BibitemShut {NoStop}%
\bibitem [{\citenamefont {Prozorov}\ and\ \citenamefont
  {Kogan}(2018)}]{Prozorov2018}%
  \BibitemOpen
  \bibfield  {author} {\bibinfo {author} {\bibfnamefont {R.}~\bibnamefont
  {Prozorov}}\ and\ \bibinfo {author} {\bibfnamefont {V.~G.}\ \bibnamefont
  {Kogan}},\ }\bibfield  {title} {\bibinfo {title} {Effective demagnetizing
  factors of diamagnetic samples of various shapes},\ }\href
  {https://doi.org/10.1103/PhysRevApplied.10.014030} {\bibfield  {journal}
  {\bibinfo  {journal} {Phys. Rev. Applied}\ }\textbf {\bibinfo {volume}
  {10}},\ \bibinfo {pages} {014030} (\bibinfo {year} {2018})}\BibitemShut
  {NoStop}%
\bibitem [{\citenamefont {Bean}\ and\ \citenamefont
  {Livingston}(1964)}]{Bean1964a}%
  \BibitemOpen
  \bibfield  {author} {\bibinfo {author} {\bibfnamefont {C.~P.}\ \bibnamefont
  {Bean}}\ and\ \bibinfo {author} {\bibfnamefont {J.~D.}\ \bibnamefont
  {Livingston}},\ }\bibfield  {title} {\bibinfo {title} {{S}urface {B}arrier in
  {T}ype-{II} {S}uperconductors},\ }\href
  {https://doi.org/10.1103/PhysRevLett.12.14} {\bibfield  {journal} {\bibinfo
  {journal} {Phys. Rev. Lett.}\ }\textbf {\bibinfo {volume} {12}},\ \bibinfo
  {pages} {14} (\bibinfo {year} {1964})}\BibitemShut {NoStop}%
\bibitem [{\citenamefont {Brandt}(1996)}]{Brandt1996}%
  \BibitemOpen
  \bibfield  {author} {\bibinfo {author} {\bibfnamefont {E.~H.}\ \bibnamefont
  {Brandt}},\ }\bibfield  {title} {\bibinfo {title} {Superconductors of finite
  thickness in a perpendicular magnetic field: Strips and slabs},\ }\href
  {https://doi.org/10.1103/physrevb.54.4246} {\bibfield  {journal} {\bibinfo
  {journal} {Phys. Rev. B}\ }\textbf {\bibinfo {volume} {54}},\ \bibinfo
  {pages} {4246} (\bibinfo {year} {1996})}\BibitemShut {NoStop}%
\bibitem [{\citenamefont {Prozorov}(2021)}]{Prozorov2021}%
  \BibitemOpen
  \bibfield  {author} {\bibinfo {author} {\bibfnamefont {R.}~\bibnamefont
  {Prozorov}},\ }\bibfield  {title} {\bibinfo {title} {Meissner-london
  susceptibility of superconducting right circular cylinders in an axial
  magnetic field},\ }\href {https://doi.org/10.1103/physrevapplied.16.024014}
  {\bibfield  {journal} {\bibinfo  {journal} {Phys. Rev. Applied}\ }\textbf
  {\bibinfo {volume} {16}},\ \bibinfo {pages} {024014} (\bibinfo {year}
  {2021})}\BibitemShut {NoStop}%
\end{thebibliography}

\end{document}